\documentclass[aps,prc,twocolumn,showpacs,floatfix,nofootinbib,preprintnumbers,superscriptaddress,amsmath,amssymb]{revtex4-1}

\usepackage{graphicx}% Include figure files
\usepackage{dcolumn}% Align table columns on decimal point
\usepackage{bm}% bold math
\usepackage{hyperref}% add hypertext capabilities
\usepackage[mathlines]{lineno}% Enable numbering of text and display math
\usepackage{subeqnarray}
\usepackage{cases}
\usepackage{times}
\usepackage{datetime}
\usepackage{CJK}
\newcommand{\rr} {\boldsymbol{r}}

\begin{document}

\title{ Generalized Second-Order Thomas-Fermi Method for Superfluid Fermi Systems }% Force line breaks with \\

\author{J.C. Pei}
%\email{peij@pku.edu.cn}
\affiliation{State Key Laboratory of Nuclear
Physics and Technology, School of Physics, Peking University,  Beijing 100871, China}
\affiliation{State Key Laboratory of Theoretical Physics, Institute of Theoretical Physics, Chinese Academy of Sciences, Beijing 100190, China}

\author{Na Fei}
\affiliation{State Key Laboratory of Nuclear
Physics and Technology, School of Physics, Peking University,  Beijing 100871, China}

\author{Y.N. Zhang}
\affiliation{State Key Laboratory of Nuclear
Physics and Technology, School of Physics, Peking University,  Beijing 100871, China}

\author{P. Schuck}
\affiliation{Institut de Physique Nucl¨¦aire, IN2P3-CNRS, Universit¨¦ Paris-Sud, F-91406 Orsay Cedex, France}
\affiliation{Laboratoire de Physique et de Mod¨¦lisation des Milieux Condens¨¦s, CNRS et Universit¨¦ Joseph Fourier, UMR5493, 25 Avenue des Martyrs, BP 166, F-38042 Grenoble Cedex 9, France}

%\author{W. Nazarewicz}
%\affiliation{Department of Physics and Astronomy and NSCL/FRIB Laboratory, Michigan State University, East Lansing, Michigan 48824, USA}
%\affiliation{Physics Division, Oak Ridge National Laboratory, Oak Ridge, Tennessee 37831, USA}
%\affiliation{Faculty of Physics, University of Warsaw, Pasteura 5, 02-093 Warsaw, Poland}

\begin{abstract}

Using the $\hbar$-expansion of the Green's function of the Hartree-Fock-Bogoliubov equation, we extend the second-order Thomas-Fermi approximation to
generalized superfluid Fermi systems by including the density-dependent effective mass and the spin-orbit potential.
We first implement and examine the full correction terms over different energy intervals of the quasiparticle spectra in calculations
of finite nuclei. Final applications of this generalized Thomas-Fermi method are intended for various inhomogeneous superfluid Fermi systems.
\end{abstract}


%\today,
%\ampmtime\

\maketitle

%\tableofcontents

\section{Introduction}

 Semiclassical treatments of quantum systems
 are  always of  broad interests. Examples are  nuclei, metallic clusters, cold atomic gases, neutron
 star crusts, etc. This is particularly useful for large systems which challenge the capacities of supercomputers.
 In this context, the Thomas-Fermi approximation (or Local Density Approximation)
 has been extended to higher orders to include as much as possible quantum corrections.
 When pairing correlations are included, the extended Thomas-Fermi method for superconducting/superfluid systems becomes much more complicated than for normal fluid systems.
 In the past few years, only a very limited number of
 works concerning the formulism of the second-order superfluid Thomas-Fermi approach have been given in the literature~\cite{schuck,gross,csordas} with either no or very limited practical calculations~\cite{csordas}. So further studies and applications
 are very desirable.  Furthermore, the density-dependent effective mass
 and the spin-orbit potential have not been considered yet, which are essential ingredients in, e.g.,  generalized nuclear density functionals
 such as the widely used Skyrme nuclear density functionals~\cite{vautherin,PGR}.
 Also in cold Fermi gases, the spin-orbit coupling~\cite{wangpj} and the density-dependent effective mass~\cite{bulgac08}
are currently very interesting.

 The Hartree-Fock-Bogoliubov (HFB) equations~\cite{manybody}, or Bogoliubov-de Gennes equations~\cite{gennes}, have been a
 framework for superfluid Fermi systems. The first $\hbar$-expansion of the HFB density
 matrix is based on the Wigner Kirkwood transformation of the Bloch propagator~\cite{schuck}, however,  without
 practical applications.
Later, the $\hbar$-expansion of the Green's function of HFB solutions has been
 derived for superconducting systems~\cite{gross}.
  More recently, a third paper based on the Green's function method appeared with some applications to cold atomic systems~\cite{csordas}.
 We also noticed that the coarse graining treatment in superfluid Fermi gases~\cite{simonucci}
 has a connection to the second-order Thomas-Fermi method.

 Our objective of the present work is to extend the formulism to include the density-dependent effective mass and
 the spin-orbit effect together with the second-order Thomas Fermi approximation for superfluid systems.
 Although such extensions~\cite{brack} and even tensor interactions~\cite{bartel} have been achieved  long time ago for non-superfluid systems, they are not
 yet elaborated for superfluid systems.
    In this work, we first derive the full correction terms based
 on the $\hbar$-expansion of the Green's function corresponding to the HFB equations.
 In addition, we implement these second-order correction terms in  numerical examinations and
  compare the results with fully self-consistent Skyrme HFB calculations of nuclei where
 the density-dependent effective mass and the spin-orbit potential are included.
 In the past the Thomas-Fermi approximation with Gogny force has already been applied to superfluid nuclei in~\cite{kucharek1,kucharek2}.

 Recently, the adaptive multi-resolution 3D coordinate-space HFB method has been developed for complex superfluid
 systems but the computation of continuum states is still very costly~\cite{Pei14}.
 The coordinate-space HFB calculations are very useful for describing weakly-bound systems and complex-shaped systems~\cite{Pei14},
 in contrast to the conventional HFB approaches based on harmonic-oscillator basis expansions.
 A promising way to address large systems is to adopt the hybrid
HFB calculations~\cite{pei2011}, i.e., using the Thomas-Fermi approximation for high-energy quasi-continuum states and the discretized HFB solutions for low-energy states.
The hybrid HFB has first been  adopted with the zeroth-order Thomas-Fermi
 method for cold atomic systems~\cite{reidl,xjliu}.
 In this respect, large superfluid systems such as the trapped cold atoms with $10^{5-6}$ particles~\cite{partridge,shin} and
 exotic neutron star crusts in large 3D cells~\cite{mazy,stone} are numerically very challenging. In such inhomogeneous systems, the
 violation of zeroth-order Thomas-Fermi approximation related to finite-size effects and complex spatial topologies could be non-negligible~\cite{Sensarma}.

\section{Formulation}

The HFB equation in the coordinate-space representation takes the form~\cite{Pei08}:
\begin{equation}\label{hfb1}
\left[
  \begin{array}{cc}
    h(\rr)& \Delta(\rr) \\
    \Delta^{*}(\rr) & -h(\rr) \\
  \end{array}
\right]
  \left[
    \begin{array}{c}
      U_{k}(\rr) \\
      V_{k}(\rr) \\
    \end{array}
  \right]
  = E_k
    \left[
    \begin{array}{c}
      U_{k}(\rr) \\
      V_{k}(\rr) \\
    \end{array}
  \right],
\end{equation}
where $h=h_{\rm HF}-\lambda$; $h_{\rm HF}$ is the Hartree-Fock Hamiltonian; $\lambda$ is the Fermi energy (or chemical potential); $\Delta$ is the pairing potential;
$U_k$ and $V_k$ are the upper and lower components of quasi-particle
wave functions, respectively; $E_k$ is the quasi-particle energy.

The theoretical derivation of the second-order superfluid Thomas-Fermi approximation starts with the
solution of the HFB equation using the Green's function method (or Gorkov equation)~\cite{gross, csordas},

\begin{equation}
\left[ \omega \textbf{I}-
\Big(\begin{array}{cc}%
h(\rr) &\Delta(\rr)  \vspace{3pt}\\
\Delta^{*}(\rr) & -h(\rr)
\end{array}\Big)
\right]\textbf{G}(\rr,\rr',\omega)=\hbar \textbf{I}\delta(\rr-\rr').
\end{equation}

The Wigner transformation of the Green's function form of the HFB equation can be written as:
\begin{equation}
\displaystyle\Lambda [(\omega \textbf{I}-\textbf{H}(\textbf{R},\textbf{p})), \textbf{G}(\textbf{R}',\textbf{p}',\omega)]=\hbar \textbf{I},
\label{HFBe}
\end{equation}
where ${\rm \textbf{H}}$ refers to the HFB Hamiltonian, and the operator $\Lambda$ can be expanded in powers of $\hbar$:
\begin{equation}
\Lambda=\sum_{j=0}^{\infty}\hbar^j\Lambda_j=\sum_{j=0}^{\infty}\frac{\hbar^j}{j!}(\frac{-i}{2})^j\Big[(\nabla_{\textbf{R}}\cdot\nabla_{\textbf{p}'}
-\nabla_{\textbf{p}}\cdot\nabla_{\textbf{R}'})\Big]^j.
\label{operator}
\end{equation}

Similar to non-superfluid systems, the spin-orbit potential can be included as a higher-order term in terms of $\hbar$ in the Hamiltonian~\cite{grammaticos},
\begin{equation}
\textbf{H}(\textbf{R},\textbf{p})=\textbf{H}_c(\textbf{R},\textbf{p})+\textbf{H}_{so}(\textbf{R},\textbf{p}),
\end{equation}
in which ${\rm \textbf{H}}_c$ and ${\rm \textbf{H}}_{so}$ denote the central term and the spin-orbit term of the HFB Hamiltonian, respectively.
The spin-orbit Hamiltonian is defined as
\begin{equation}
 \textbf{H}_{so}=
\Big(\begin{array}{cc}%
 h_{so} & 0 \vspace{3pt}\\
0 & - h_{so}
\end{array}\Big)
, \hspace{8pt} h_{so} = -\frac{1}{\hbar}\textbf{w}(\rr)\cdot \mathbf{\sigma} \times \mathbf{p}
\end{equation}

The extended Thomas-Fermi method consists in expanding the Green's function in Eq.(\ref{HFBe}) in powers of $\hbar$ , i.e.,
\begin{equation}
\textbf{G}(\textbf{R},\textbf{p},\omega)=\hbar\sum_{n=0}^{\infty}\textbf{G}_n(\textbf{R},\textbf{p},\omega)\hbar^n .
\end{equation}

The expansion of Eq.(\ref{HFBe}) in terms of $\hbar$ to the second-order can be written as,
\begin{subequations}
\begin{equation}
\begin{array}{l}
(\omega \textbf{I}-\textbf{H}_c)\textbf{G}_0=\textbf{I},
\end{array}
\end{equation}
\begin{equation}
\begin{array}{ll}
(\omega \textbf{I}-\textbf{H}_c)\textbf{G}_1&+\Lambda_1[(\omega \textbf{I}-\textbf{H}_c), \textbf{G}_0] \vspace{5pt}\\
&+\Lambda_0[-\textbf{H}_{so}, \textbf{G}_0]=0,\\
 \end{array}
\end{equation}
\begin{equation}
\begin{array}{ll}
 (\omega \textbf{I}-\textbf{H}_c)\textbf{G}_2&+\Lambda_2[(\omega \textbf{I}-\textbf{H}_c), \textbf{G}_0] \vspace{5pt}\\
 &+ \Lambda_1[(\omega \textbf{I}-\textbf{H}_c), \textbf{G}_1] \vspace{5pt}\\
  & +\Lambda_1[-\textbf{H}_{so}, \textbf{G}_0]\vspace{5pt}\\
 & +\Lambda_0[-\textbf{H}_{so}, \textbf{G}_1]=0 .\\
 \end{array}
\end{equation}
\label{HFBe2}
\end{subequations}

The various expansion terms of $\textbf{G}$ can be obtained order by order.
Compared to Ref.\cite{gross}, we include the spin-orbit potential entailing more terms to appear in Eq.(\ref{HFBe2}).
First, $\textbf{G}_0$ corresponds to the
zeroth-order solutions and does not change due to the spin-orbit potential.
The matrix elements of $\textbf{G}_0$ are written as,
\begin{equation}
\Big(\begin{array}{cc}%
\textbf{G}_0^{(\uparrow\uparrow)} & \textbf{G}_0^{(\uparrow\downarrow)} \vspace{3pt}\\
\textbf{G}_0^{(\downarrow\uparrow)} & \textbf{G}_0^{(\downarrow\downarrow)}
\end{array}\Big)=\frac{1}{D}
\Big(\begin{array}{cc}%
\omega+h & \Delta \vspace{3pt}\\
\Delta & \omega-h
\end{array}\Big),
\label{hfbe3}
\end{equation}
where the $\textbf{G}_0^{(\uparrow\uparrow)}$ and $\textbf{G}_0^{(\uparrow\downarrow)}$
correspond to the normal density and pairing density, respectively. The abbreviation $D$ for the denominator in Eq.(\ref{hfbe3})
is defined as $D=(\omega^{2}-h^{2}-\Delta^2)$.

$\textbf{G}_1$  changes due to the spin-orbit potential and is written as
\begin{equation}
\begin{split}
&\textbf{G}_{1,\textbf{so}}^{(\uparrow\uparrow)}=\frac{1}{D^2} h_{so}[(\omega+h)^2-\Delta^2] \vspace{5pt} \\
&\textbf{G}_{1,\textbf{so}}^{(\uparrow\downarrow)} = \frac{1}{D^2}h_{so}\big( 2h\Delta \big)\\
\end{split}
\end{equation}
Note that in this work we are treating systems in the case of real pairing potentials. The formulism with complex pairing
potentials can be derived similarly but would be
more complicated~\cite{gross}.

$\textbf{G}_2$ is quite complicated~\cite{gross,csordas} and the additional terms due to the spin-orbit potential are:
\begin{equation}
\begin{split}
&\displaystyle\textbf{G}_{2,\textbf{so}}^{(\uparrow\uparrow)}=\displaystyle\frac{h_{so}^2}{D^3}\big[(\omega+h)^3-(\omega+3h)\Delta^2\big]
\vspace{5pt}\\\vspace{5pt}
&\displaystyle\textbf{G}_{2,\textbf{so}}^{(\uparrow\downarrow)}=\displaystyle\frac{h_{so}^2}{D^3}\big[(\omega^2-\Delta^2+3h^2)\Delta\big]
\end{split}
\end{equation}
Note that terms involving derivatives of the spin-orbit potential have been omitted.
Actually the terms in $\textbf{G}_2$ with $h_{so}$ do not contribute and only terms with $h_{so}^2$ do.

Next, the contributions to normal density and pairing density can be obtained by an appropriate integration of $\textbf{G}_2$ in
the complex $\omega$  plane.
For convenience of writing the expressions, we introduce a differential operator $\overleftrightarrow{\Lambda}$ as in Ref.~\cite{schuck},
\begin{equation}
\overleftrightarrow{\Lambda}=(\overleftarrow{\nabla}_{\textbf{R}}\overrightarrow{\nabla}_{\textbf{p}}-\overleftarrow{\nabla}_{\textbf{p}}\overrightarrow{\nabla}_{\textbf{R}}),
\label{poisson}
\end{equation}
which is the operator in the  Poisson bracket. Eq.(\ref{poisson}) can be applied repeatedly and is related to Eq.(\ref{operator}) without expansion coefficients.
 The 2$^{\rm nd}$-order normal density contribution terms are given as,
\begin{equation}
\begin{array}{lcl}
\rho_2(\textbf{R},\textbf{p})&=&\eta_1(h\overleftrightarrow{\Lambda}^2h)+\eta_2(h\overleftrightarrow{\Lambda}^2\Delta)+\eta_3(\Delta\overleftrightarrow{\Lambda}^2\Delta) \vspace{5pt}\\
&& +\eta_4(h(\overleftrightarrow{\Lambda} h)^2)+\eta_5(h(\overleftrightarrow{\Lambda} h)(\overleftrightarrow{\Lambda} \Delta))\vspace{5pt}\\
&& + \eta_6(h(\overleftrightarrow{\Lambda} \Delta)^2)+\eta_7(\Delta(\overleftrightarrow{\Lambda} h)^2)\vspace{5pt}\\
&& + \eta_8 (\Delta(\overleftrightarrow{\Lambda} h)(\overleftrightarrow{\Lambda}\Delta))+\eta_9(\Delta(\overleftrightarrow{\Lambda}\Delta)^2)\vspace{5pt}\\
&& + \eta_{10} (h\overleftrightarrow{\Lambda}\Delta)^2+\displaystyle\frac{3h\Delta^2h_{so}^2}{4E^5}
\end{array}
\label{ndensity}
\end{equation}
and the 2$^{\rm nd}$-order pairing density,
\begin{equation}
\begin{array}{lcl}
\tilde{\rho}_2(\textbf{R},\textbf{p})&=&\theta_1(h\overleftrightarrow{\Lambda}^2h)+\theta_2(h\overleftrightarrow{\Lambda}^2\Delta)+\theta_3(\Delta\overleftrightarrow{\Lambda}^2\Delta)\vspace{5pt} \\
&& +\theta_4(h(\overleftrightarrow{\Lambda }h)^2)+\theta_5(h(\overleftrightarrow{\Lambda} h)(\overleftrightarrow{\Lambda} \Delta))\vspace{5pt}\\
&& + \theta_6(h(\overleftrightarrow{\Lambda} \Delta)^2)+\theta_7(\Delta(\overleftrightarrow{\Lambda} h)^2)\vspace{5pt}\\
&& + \theta_8 (\Delta(\overleftrightarrow{\Lambda}h)(\overleftrightarrow{\Lambda}\Delta))+\theta_9(\Delta(\overleftrightarrow{\Lambda}\Delta)^2)\vspace{5pt}\\
&& + \theta_{10} (h\overleftrightarrow{\Lambda}\Delta)^2+\displaystyle\frac{(2h^2-\Delta^2)\Delta h_{so}^2}{4E^5}
\end{array}
\label{pdensity}
\end{equation}
In Eq.(\ref{ndensity}) and Eq.(\ref{pdensity}), the coefficients $\eta_i$ and $\theta_i$ have been given in Refs.~\cite{schuck,gross,csordas},
although they have different conventions. We refer to the Appendix for explicit expressions of these terms.
Note that some terms  vanish with the application of operators $\Lambda_j$, since we suppose a momentum-independent pairing potential.
The additional contributions due to the spin-orbit potential are given in Eqs.(\ref{ndensity}) and (\ref{pdensity}) explicitly, in which
$E$ denotes the HFB quasiparticle energies and takes the form as
\begin{equation}
E=\sqrt{
(h_{\rm HF}(\mathbf{R},\mathbf{p})-\lambda)^2 +\Delta(\mathbf{R},\mathbf{p})^2} .
\end{equation}

Up to now, we have not explicitly considered the density-dependent effective mass.
The density-dependent effective mass is involved in the central part of the single-particle Hamiltonian,
\begin{equation}
h_c=\frac{1}{2}f(\rr)p^2+U(\rr)~~~~~\mbox{with}~~~m^{*}(\rr)=1/f(\rr).
\end{equation}
In fact, the expressions of the $2^{\rm nd}$-order Thomas-Fermi do not need to be modified.
However, three additional expansion terms appear compared to the
cases without effective mass:
\begin{subequations}
\begin{equation}
 h\overleftrightarrow{\Lambda}^2h=\displaystyle fp^2\nabla^2f+2f\nabla^2U-\frac{2}{3}p^2(\nabla f)^2\\
 \end{equation}
 \begin{equation}
 \begin{array}{lcl}
h(\overleftrightarrow{\Lambda}h)^2&=&\displaystyle-\frac{1}{12}fp^4(\nabla f)^2+f(\nabla U)^2+\frac{1}{6}f^2p^4\nabla^2f\vspace{5pt}\\
&&\displaystyle+\frac{1}{3}f^2p^2\nabla^2U+\frac{1}{3}fp^2(\nabla f)(\nabla U) \\
\end{array}\\
\end{equation}
\begin{equation}
 h(\overleftrightarrow{\Lambda} h)(\overleftrightarrow{\Lambda} \Delta)=\displaystyle\frac{1}{6}fp^2(\nabla f)(\nabla \Delta)+f(\nabla U)(\nabla \Delta)
\end{equation}
\end{subequations}

Finally, the normal density (and pairing density) in real space can be obtained by integrals over quasiparticle energies $E$
or single-particle energies $\varepsilon$ up to a certain cutoff
\begin{equation}
\begin{array}{lcl}
\rho_2(\rr)&=&\displaystyle\int\frac{\rho_2(\textbf{R},\textbf{p})}{(2\pi \hbar)^3} d^3p \vspace{5pt}\\
&=&\displaystyle\int_{E0}^{Ec}\frac{\rho_2(\textbf{R},\textbf{p})}{(2\pi \hbar)^3} \frac{pE}{f\sqrt{E^2-\Delta^2}}dE \vspace{5pt}\\
&=&\displaystyle\int_{\varepsilon 0}^{\varepsilon c}\frac{\rho_2(\textbf{R},\textbf{p})}{(2\pi \hbar)^3} \frac{p}{f}d\varepsilon
\end{array},
\end{equation}
where $(E_0, E_c)$ or $(\varepsilon_0, \varepsilon_c)$  defines an energy interval in which the Thomas-Fermi approximation is used.

The 2$^{\rm nd}$-order kinetic density correction in real spaces is obtained as,
\begin{equation}
\tau_2(\rr)= \displaystyle \frac{1}{4}\nabla^2\rho +\int_{E0}^{Ec}\frac{\rho_2(\textbf{R},\textbf{p})}{(2\pi \hbar)^3} \frac{p^3E}{f\sqrt{E^2-\Delta^2}}dE
\end{equation}

The 2$^{\rm nd}$-order spin-orbit density correction is given as:
\begin{equation}
\textbf{J}_2(\rr)= \displaystyle \int\frac{\rho(\textbf{R},\textbf{p})}{(2\pi \hbar)^3} (\textbf{p}\times {\bf{\sigma}}) d^3p
\end{equation}
Due to the spin degeneracy, various correction densities should be multiplied by a factor of 2.
Note that there is no zeroth-order spin-orbit density. Since the liner term of $(\textbf{p}\times {\bf{\sigma}})$ doesn't contribute in the integration, the final expression
of $\textbf{J}_2(\rr)$ can be obtained by considering only the  $\textbf{G}_{1,\textbf{so}}$'s contribution,
\begin{equation}
\textbf{J}_2(\rr)= \frac{\textbf{w}(\rr)}{(\pi \hbar)^2}\displaystyle \int_{E0}^{Ec}  \frac{-\Delta^2}{3E^3}  \frac{p^3E}{f\sqrt{E^2-\Delta^2}}dE.
\end{equation}

\section{The limit of zero pairing gap}

In the limit of the zero pairing gap,  the derived generalized 2$^{\rm nd}$-order Thomas-Fermi approximation for the
normal state should be recovered. For the normal density and pairing density, this has been demonstrated in Ref.~\cite{gross}.
In this work, we have to examine the additional terms of spin-orbit contributions and the spin-orbit density $\textbf{J}(\rr)$ in the
limit of zero pairing gap. For the spin-orbit contribution to normal density:
\begin{equation}
\lim_{\Delta \to 0} \displaystyle\frac{3h\Delta^2h_{so}^2}{4E^5} = -\frac{1}{2}h_{so}^2\delta'(h).
\end{equation}
The resulting density expression is equivalent to the 2$^{\rm nd}$-order Thomas-Fermi approximation of the non-superfluid state~\cite{grammaticos}.
Obviously, the resulting pairing density due to the spin-orbit contribution becomes zero in the limit of the zero pairing gap.
Similarly, the spin-orbit density in the limit can be obtained,
\begin{equation}
\begin{array}{ccl}
\lim\limits_{\Delta \to 0} \textbf{J}_2(\rr) &= &\displaystyle-2\frac{\textbf{w}(\rr)}{(\pi \hbar)^2} \displaystyle \int_{\varepsilon0}^{\varepsilon c}  \frac{\delta(h)}{3} \frac{p^3}{f}d\varepsilon \vspace{5pt} \\
&=&\displaystyle -2\rho(\rr)\frac{\textbf{w}(\rr)}{f(\rr)}\\
\end{array}
\end{equation}
which also agrees with the expression of the normal state~\cite{grammaticos}.

\section{Implementation and calculations}

\begin{figure}[t]
  \includegraphics[width=0.48\textwidth]{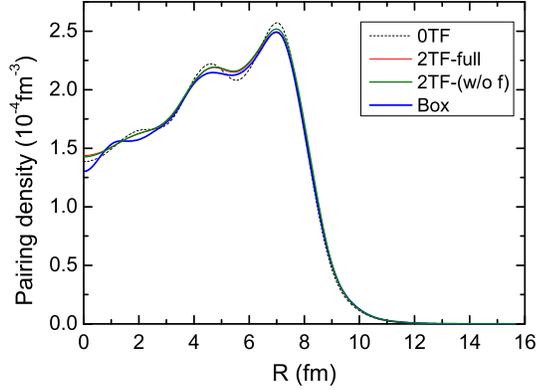}\\
  \caption{(Color online) The 0$^{\rm th}$-order (dotted line) and 2$^{\rm nd}$-order Thomas-Fermi contributions (0$^{\rm th}$-order plus 2$^{\rm nd}$-order ) to the neutron pairing density
  from 55 to 65 MeV, as well as the corresponding coordinate-space HFB solutions (as labeled by `Box'). The 2$^{\rm nd}$-order Thomas-Fermi contribution
  without explicitly considering the effective mass is also shown, as labeled by `2TF-(w/o f)'.  }
  \label{fig-pdensity}
\end{figure}

 \begin{figure}[t]
  \includegraphics[width=0.48\textwidth]{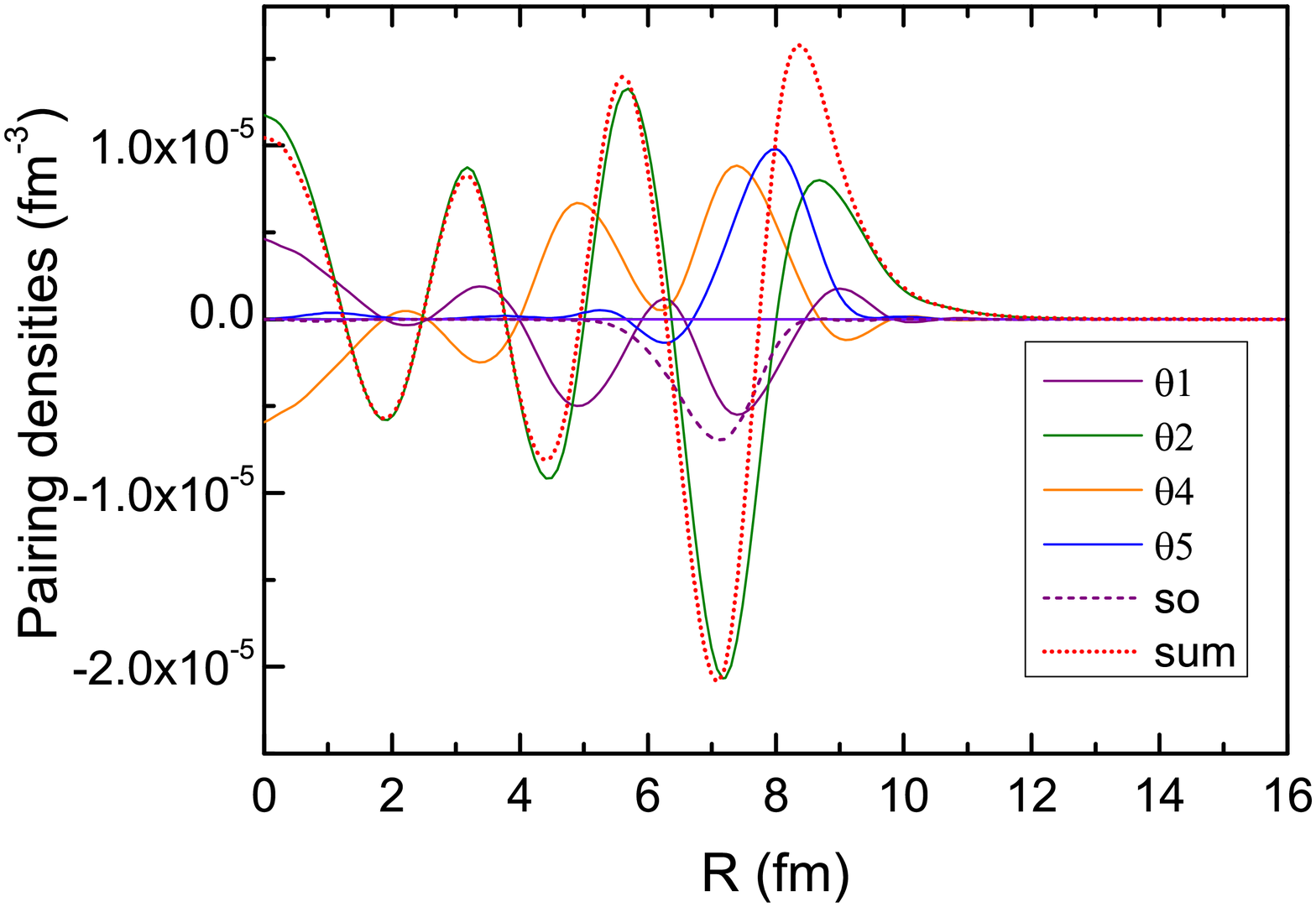}\\
  \caption{(Color online) The 2$^{\rm nd}$-order Thomas-Fermi terms corresponding to the pairing density in Fig.\ref{fig-pdensity}. The terms denoted by $\theta_i$
  as listed in Eq.(\ref{pdensity}). The spin-orbit contribution (dashed line) and the summed correction (dotted line) are shown.  }
  \label{fig-pdensity-t}
\end{figure}

 \begin{figure}[t]
  \includegraphics[width=0.48\textwidth]{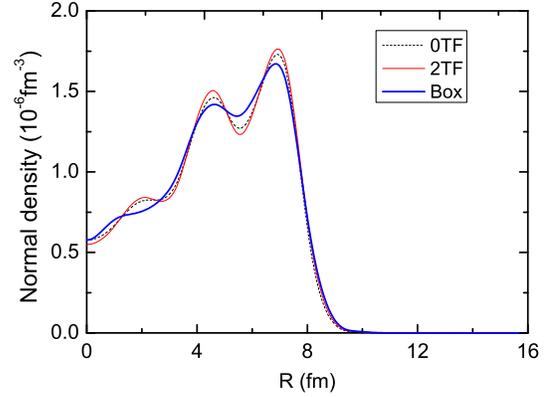}\\
  \caption{(Color online) Similar to Fig.\ref{fig-pdensity}, but for the 0$^{\rm th}$-order and 2$^{\rm nd}$-order Thomas-Fermi contributions to the neutron normal density. The
  normal density from self-consistent coordinate-space HFB solutions is labeled by `Box'.  }
  \label{fig-density}
\end{figure}

 \begin{figure}[t]
  \includegraphics[width=0.48\textwidth]{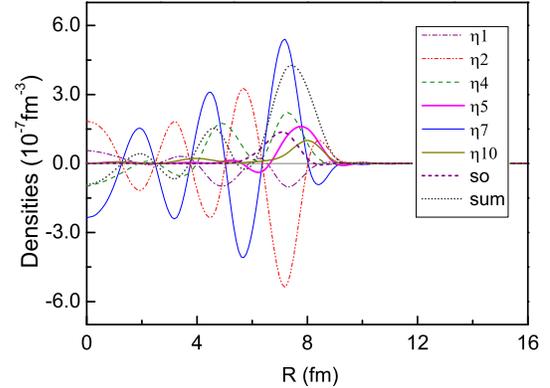}\\
  \caption{(Color online) Similar to Fig.\ref{fig-pdensity-t}, but for the 2$^{\rm nd}$-order contribution terms to the normal density as listed in Eq.(\ref{ndensity}). }
  \label{fig-density-t}
\end{figure}

 \begin{figure}[t]
  \includegraphics[width=0.48\textwidth]{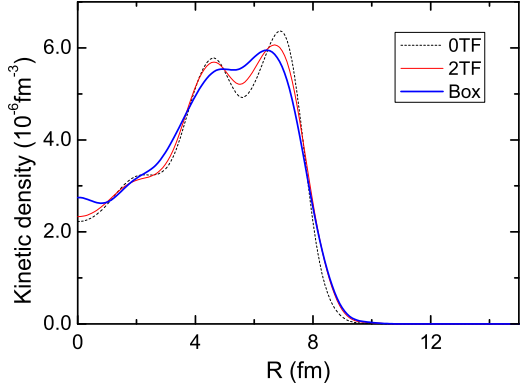}\\
  \caption{(Color online) Similar to Fig.\ref{fig-pdensity}, but for the 0$^{\rm th}$-order and 2$^{\rm nd}$-order Thomas-Fermi contributions to the neutron  kinetic density. }
  \label{fig-kdensity}
\end{figure}

 \begin{figure}[t]
  \includegraphics[width=0.48\textwidth]{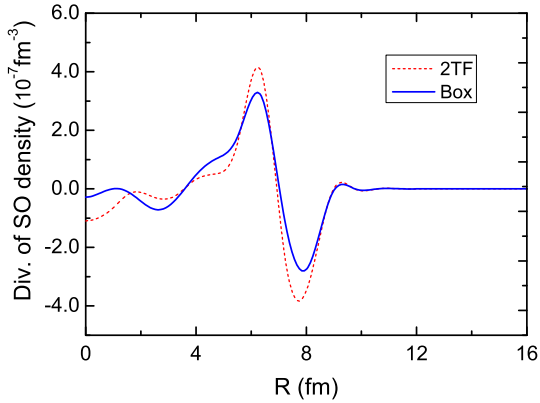}\\
  \caption{(Color online) Similar to Fig.\ref{fig-pdensity}, but for the  2$^{\rm nd}$-order Thomas-Fermi  contribution (0$^{\rm th}$-order doesn't contribute) to the divergence of the neutron spin-orbit density $\nabla\cdot \textbf{J}$. }
  \label{fig-sodensity}
\end{figure}

 \begin{figure}[t]
  \includegraphics[width=0.45\textwidth]{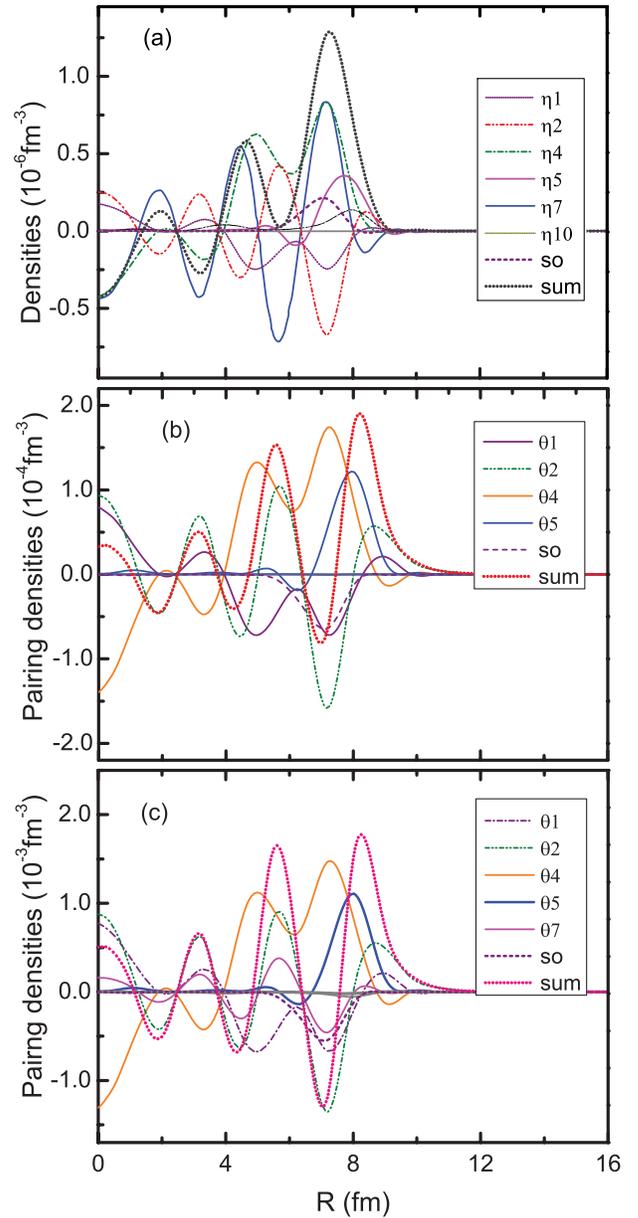}\\
  \caption{(Color online) (a) The 2$^{\rm nd}$-order Thomas-Fermi terms to the neutron normal density with an energy cutoff from 25 MeV to 65 MeV; (b) the 2$^{\rm nd}$-order terms
  to the pairing density; (c) the 2$^{\rm nd}$-order terms to the pairing density
  with a pairing strength 10 times larger. }
  \label{fig-density-25}
\end{figure}

 \begin{figure}[t]
  \includegraphics[width=0.48\textwidth]{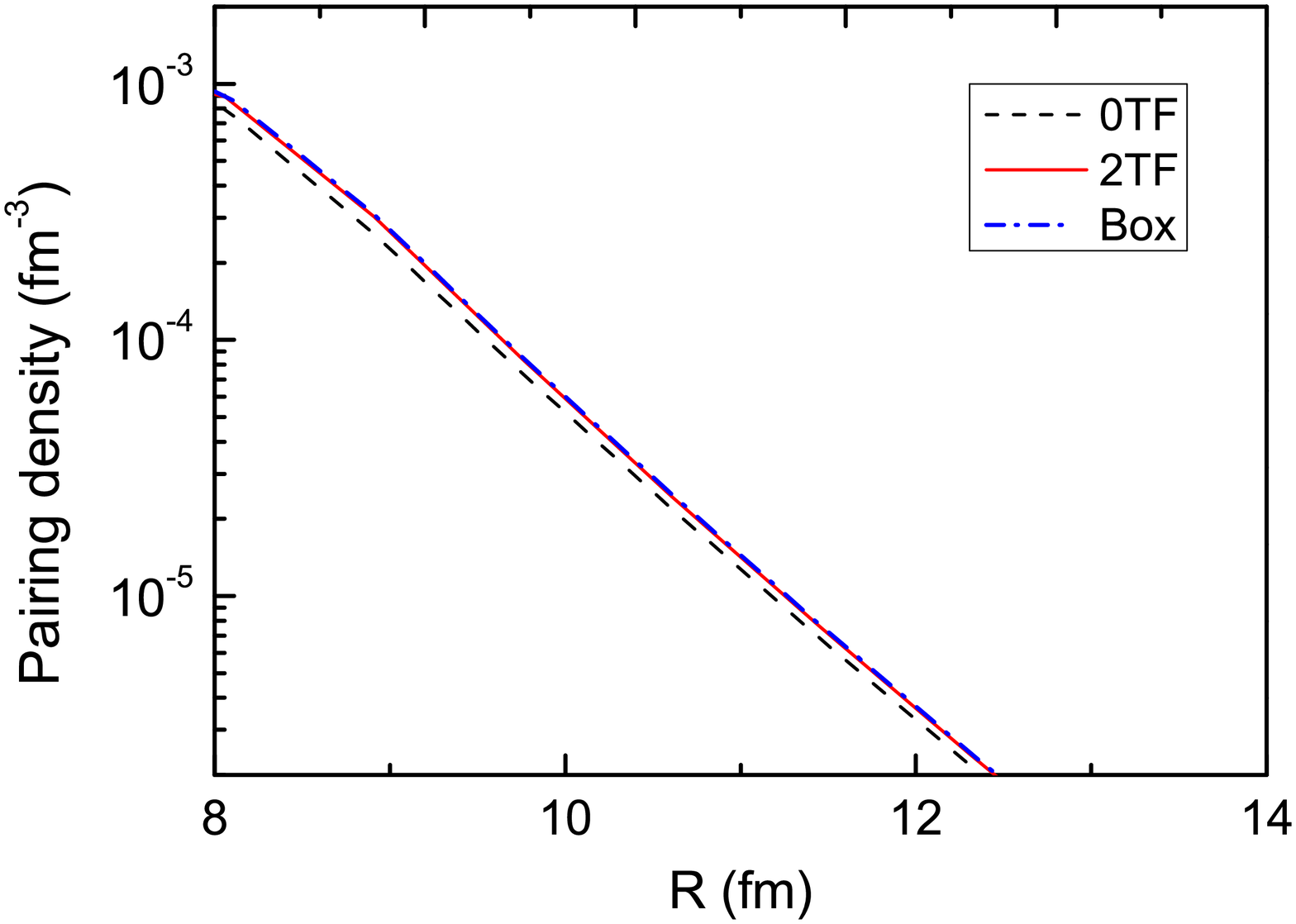}\\
  \caption{(Color online) The neutron pairing density corresponding to Fig.\ref{fig-density-25}(c) plotted on a logarithmic scale to show surface distributions, compared
  to the coordinate-space HFB solutions. }
  \label{fig-psur}
\end{figure}

To demonstrate applications of the 2$^{\rm nd}$-order Thomas-Fermi corrections, we perform self-consistent calculations of
a finite deformed nucleus $^{^{238}}$U.
For the particle-hole interaction channel, the often used Skyrme force SLy4~\cite{sly4} is adopted.
For the pairing channel, the volume pairing, i.e., a delta interaction for the pairing force is adopted with
a reasonable pairing strength of 200 MeV fm$^{-3}$.
The Skyrme density functional contains the density-dependent effective mass
and the spin-orbit potential, providing an ideal testing ground for the generalized 2$^{\rm nd}$-order Thomas-Fermi approximation.

\subsection{Non-self-consistent Calculations}

The nuclear HFB solutions have deep-hole states~\cite{pei2011} which correspond to deep-bound states in Hartree-Fock+BCS solutions.
In the HFB approach, deep-hole states are narrow quasiparticle resonances with large quasiparticle energies $E_k$ and large occupation numbers $v_k^2$.
These states contain important shell effects and can not be well described by
the Thomas-Fermi approximation, since the latter does not account for shell effects. Therefore, we first aim to compare the contributions of various partial densities corresponding to
high quasiparticle energies. Our final goal is to combine the HFB solutions at low-energy quasiparticle
states and 2$^{\rm nd}$-order corrections in the quasiparticle high-energy  region.
In such a hybrid way, the computational costs for large superfluid Fermi systems can be remarkably reduced.
Indeed, the high-energy states behave like quasi-continuum states and are distinctly different from low-energy states.
One can suppose that the former can be well approximated by the Thomas-Fermi expressions.

First, self-consistent HFB calculations are performed for $^{^{238}}$U, then the zeroth-order and second-order
Thomas-Fermi approximations are investigated by performing one iteration with densities from full HFB solutions.
 The HFB calculations are performed in the cylindrical coordinate space
with the HFB-AX solver~\cite{Pei08} that uses B-spline techniques.
For calculations employing large box sizes and small lattice spacings,
the discretized continuum spectra would be very dense, providing good resolutions.
 The pairing densities $\tilde{\rho}(r,z)$ are plotted along the diagonal
coordinate $R=\sqrt{r^2+z^2}$$|_{(r=z)}$ rather than the axes, to avoid numerical errors at boundaries.
We compare various pairing densities in Fig.\ref{fig-pdensity}.
In this work, our discussions are restricted to neutrons since pairing
is absent in protons in $^{^{238}}$U.
The pairing densities of the full self-consistent HFB calculation are summed for quasiparticle energies ranging from 55 to 65 MeV. This is the high energy window for which we want to study the Thomas-Fermi approximation, in discarding the influence of deep-hole states. The upper cutoff is taken as 65 MeV in all our results. The converged HFB densities are then inserted into the Thomas-Fermi expressions
for one iteration and
the zeroth and second order Thomas-Fermi densities are also summed (integrated) over the same energy interval.
In this way the quantities shown in Fig.\ref{fig-pdensity} are perturbative.
It can be seen that the second-order contribution is very small
but can slightly improve the pairing density distribution.
Indeed, as shown in Eq.(\ref{ndensity}) and Eq.(\ref{pdensity}), the  second-order corrections involve terms of order $1/E^5$ and $1/E^7$, which
are suppressed in the high-energy region.  With the derivative terms of effective mass, the second-order corrections are only slightly modified in the high
energy region.

In Fig.\ref{fig-pdensity-t}, the second-order contributions from different terms in Eq. (\ref{pdensity}) are displayed corresponding to Fig.\ref{fig-pdensity}.
As we can see, the spin-orbit effect has non-negligible contributions. The term of $h\overleftrightarrow{\Lambda}^2\Delta$ is dominating
in the second-order pairing density.  In the coarse graining treatment of Bogoliubov-de Gennes equations, only this term has been included in the pairing equations ~\cite{simonucci}.
Based on our full analysis, we see that such a coarse graining treatment is reasonable, in particular at the surface.
The terms of $h(\overleftrightarrow{\Lambda} h)^2$ and $h\overleftrightarrow{\Lambda}^2h$ are non-negligible but the two
almost cancel each other in the high energy region.

Similar to Fig.\ref{fig-pdensity}, the contributions to normal densities are displayed in Fig.\ref{fig-density}. Compared to Fig.\ref{fig-pdensity},  the high-energy contributions to the normal density are much smaller (by 2 orders of magnitude) than
to the pairing density.
The second-order correction can improve the density distribution at the surface. However, the second-order correction
 leads to enhanced oscillations in the inner part of the density distribution, due to non-self-consistent calculations.
  In Fig.\ref{fig-density-t}, we see that the terms of $\Delta(\overleftrightarrow{\Lambda}h)^2$
and $h\overleftrightarrow{\Lambda}^2 \Delta$ are dominating but the two almost cancel each other. The terms of $h\overleftrightarrow{\Lambda}^2h$
and $h(\overleftrightarrow{\Lambda}h)^2$ are also canceling each other.
The term of $h(\overleftrightarrow{\Lambda}\Delta)^2$ is negligible. In contrast to the pairing density, it is hard to say which term plays a major role in the normal density.

In Fig.\ref{fig-kdensity}, the corrections to kinetic densities are displayed. Again, the second-order correction improves the description at surfaces  compared
to the zero-order correction. In addition, the second-order correction also improves the kinetic density in the inner part.
In Fig.\ref{fig-sodensity}, the second-order correction to the divergence of the spin-orbit density, $\nabla\cdot \textbf{J}$, is shown. Note that there is no zero-order correction to
the spin-orbit density. The obtained $\nabla\cdot \textbf{J}$ roughly agrees with the HFB solutions. It is important to obtain the spin-orbit density
to do self-consistent semiclassical calculations.

 To study the cutoff dependence, the second-order Thomas-Fermi correction terms to the densities with
an enlarged energy interval from 25 MeV to 65 MeV are shown in Fig.\ref{fig-density-25}.
In Fig.\ref{fig-density-25}(a), the corrections to normal densities are displayed.  Compared to
Fig.\ref{fig-density-t}, the resulting densities are increased by a factor of 20. It is obvious that the terms of $h\overleftrightarrow{\Lambda}^2h$ and $h(\overleftrightarrow{\Lambda}h)^2$ do not cancel each other anymore.
The same holds for the terms of $\Delta(\overleftrightarrow{\Lambda}h)^2$
and $h\overleftrightarrow{\Lambda}^2 \Delta$. The term  $h(\overleftrightarrow{\Lambda}\Delta)^2$ is still negligible.
In Fig.\ref{fig-density-25}(b), the corrections to pairing densities are shown. Compared to
Fig.\ref{fig-pdensity-t}, the resulting pairing densities are increased by a factor of 10. In this case, we see that
the total correction do not follow the term $h\overleftrightarrow{\Lambda}^2\Delta$ anymore which was dominating in the high energy window.
The terms of $h(\overleftrightarrow{\Lambda} h)^2$ and $h\overleftrightarrow{\Lambda}^2h$ do not
cancel each other. Or, in other words, the coarse graining treatment is not applicable in the low-energy region. The terms of
second-order have to be fully
taken into account for low-energy quasiparticle states.

In fact, the pairing in nuclei is rather weak, compared to cold atomic systems, for instance,  close to the unitary limit.
Therefore it is instructive to study for what happens if we increase artificially the pairing.  In Fig.\ref{fig-density-25}(c), the resulting corrections to the pairing densities are obtained with a 10-times larger pairing strength compared
to Fig.\ref{fig-density-25}(b). In this case, the term of $\Delta(\overleftrightarrow{\Lambda} h)^2$ acquires a role which is absent in Fig.\ref{fig-density-25}(b).

In Fig.\ref{fig-psur}, the pairing density contribution corresponding to Fig.\ref{fig-density-25}(c) are shown on a log scale.
It can be seen that the second-order correction agrees exactly with the asympotics of coordinate-space HFB solutions at large distances.
While the zeroth-order correction underestimates the pairing density  at the nuclear surface by 10\%.
These examples demonstrate clearly the advantages of the full second-order Thomas-Fermi approximation for quasiparticle energy intervals with a lower edge reaching into the low energy domain (e.g., 25 MeV).

\subsection{Hybrid HFB Calculations}

 \begin{figure}[t]
  \includegraphics[width=0.48\textwidth]{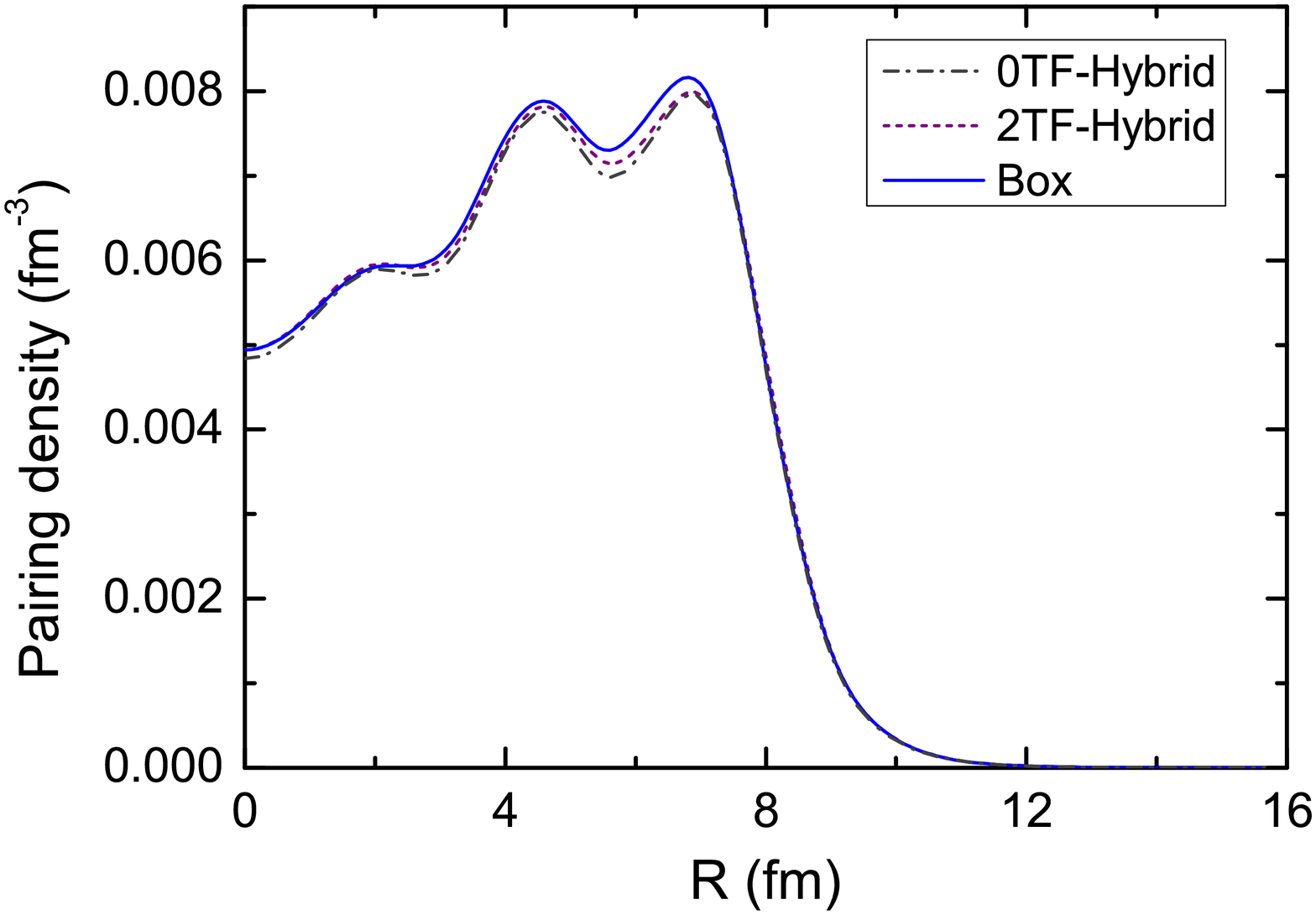}\\
  \caption{(Color online) The hybrid HFB calculations of neutron pairing density in $^{238}$U with 0$^{\rm th}$-order and 2$^{\rm nd}$-order Thomas-Fermi approximations within an quasiparticle energy interval from 25 MeV to 65 MeV, compared to the HFB-AX calculations.  }
  \label{fig-hpairing}
\end{figure}

 \begin{figure}[t]
  \includegraphics[width=0.48\textwidth]{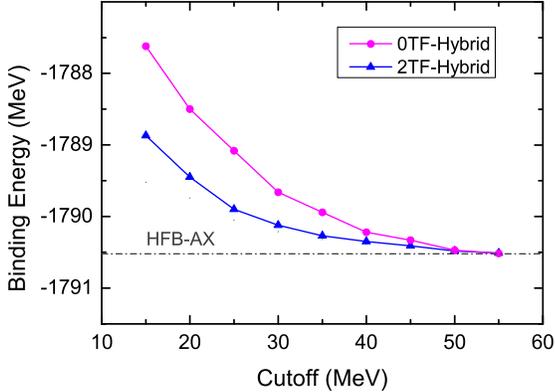}\\
  \caption{(Color online) The hybrid HFB calculations of binding energies of $^{238}$U with 0$^{\rm th}$-order and 2$^{\rm nd}$-order Thomas-Fermi approximations at different
  cutoffs, compared to the HFB-AX calculations.  }
  \label{fig-cutoff}
\end{figure}

One of our main motivations to implement the hybrid HFB calculations with the second-order Thomas-Fermi approximation is
to reduce the computing costs for large systems. In our previous work, we have tested the
hybrid HFB strategy with the zeroth-order Thomas-Fermi approximation~\cite{pei2011}. In the hybrid strategy,
the high energy deep-hole states and continuum states are separately treated. In the present
work, the high-energy continuum is treated in the second-order Thomas-Fermi approximation.
The deep-hole states, which are narrow quasiparticle resonances, can be described separately.
Firstly, we diagonalize the single-particle Hamiltonian to obtain the deep-bound Hartree-Fock wavefunctions and single-particle energies.
\begin{subequations}
\begin{equation}
(h-\lambda)v_i^{HF}(\rr)= \varepsilon_i v_i^{HF}(\rr) \vspace{5pt}
\end{equation}
\begin{equation}
(h-\lambda+\varepsilon_i)u_i^{HF}(\rr)=\Delta(\rr) v_i^{HF}(\rr)
\label{HF2}
\end{equation}
\label{H-HFB}
\end{subequations}
Next, one natural way is to renormalize the deep-hole wavefunctions with the BCS approximation, as in Ref.~\cite{pei2011}.
We solve the Eq.(\ref{HF2}) to describe the asympotics of scattering components in stead of BCS.
Another more elaborate way is to  solve the HFB equation perturbatively, as described in Ref.~\cite{bulgac}.

In the hybrid strategy, we introduce an energy interval (or window) of  quasiparticle energies from a lower cutoff to 65 MeV. Below the lower cutoff, the
box discretization is used to solve the HFB equation using the HFB-AX solver. Within the window, the zeroth-order or second-order Thomas-Fermi  approximation
for the continuum and the BCS approximation for deep-hole states are used. The quantal effects are taken into account by the standard HFB solutions below lower cutoffs
and the deep-hole states.
The hybrid HFB calculations are
implemented self-consistently to keep the conservation of particle numbers. With iterations in terms of various densities, i.e.,
the normal density $\rho(\rr)$, the kinetic density $\tau(\rr)$, the spin-orbit density $\textbf{J}(\rr)$ (does not appear in the zeroth-order),
the pairing density $\tilde{\rho}(\rr)$, the self-consistent
hybrid Skyrme HFB calculations can be realized.
The quantal and semiclassical are coupled via the summed densities.
In our test calculations of $^{238}$U,
we use the SLy4 force and the volume pairing interaction, as we used in the non-self-consistent calculations.
In Fig.\ref{fig-hpairing}, the total pairing density profiles from hybrid calculations are shown.
It can be seen that the agreements between hybrid and full HFB calculations are quite good. \\

 Concerning the treatment of the deep hole states, we should note that the BCS underestimates the pairing correlation
compared to the HFB approach, due to the absent of continuum coupling. For example, to reproduce
the neutron pairing gap of 1.245 MeV in $^{120}$Sn with SLy4 and the volume pairing, the pairing strengthes in BCS and HFB
have to be adjusted to 283 and 187 MeV fm$^{-3}$, respectively.
Hence we slightly increase the pairing in the BCS treatment of deep-hole states with one global factor.
 As a result the average pairing gaps at all cutoffs are close to the full HFB result.
For systems such as trapped Fermi gases, there are no deep-hole states and the hybrid calculations will be simpler.

The results of hybrid HFB with different lower cutoffs of the energy windows
are displayed in Fig.\ref{fig-cutoff}. In Fig.\ref{fig-cutoff}, the deviations in total binding energies between the hybrid HFB and the coordinate-space HFB solutions increase as the lower cutoff decreases.
Due to the self-consistency, the second-order  approximation  is obviously better than the zero-order Thomas-Fermi, in particular at low-energy cutoffs.
As we can see, with the cutoff energy at 25 MeV, the deviation is $\thicksim$0.5 MeV over the total binding energy of 1790.52 MeV, which is quite satisfactory.
Note that the deviations are not only from the approximate treatment of continuum but also from the treatment of
deep-hole states. In the low-energy region, quasiparticle resonances can acquire considerable widths, leading to ambiguous contaminations.
In addition, the numerical accuracy of derivatives is important in the second-order calculations. It
  will be improved by the multi-wavelet techniques~\cite{Pei14} in the future.
The full 3D coordinate-space HFB calculations are very expensive even with the efficient multi-wavelet techniques~\cite{Pei14}.
 With the lower cutoff of 25 MeV,  the number of eigen-functions to be solved would be reduced by half.
On the other hand the extra numerical cost to include the second order Thomas-Fermi term with respect to take only the zeroth-order, is almost negligible.
In this case, the resulting computational cost can be reduced at least by one order of magnitude and the desired accuracy
is still retained.

\section{Summary}

 In summary, we extended the second-order Thomas-Fermi approximation of the Hartree-Fock-Bogoliubov solutions
 for superfluid systems by including the effective mass and the spin-orbit potential, which are essential for full calculations.
 In particular, the spin-orbit contribution is important because it is absent in the zeroth-order Thomas-Fermi approximation.
 The expressions of the  new terms have been checked in the zero-pairing limit.
 The full second-order terms
 have been examined numerically in a perturbative manner in comparison with self-consistent coordinate-space HFB  solutions.
 In general, the second-order corrections can improve various density distributions at surfaces
 compared to the zeroth-order corrections. The significance of including full superfluid second-order corrections
 definitely increases as the lower edge of the  energy interval in which the Thomas-Fermi approximation acts goes down.
 Among the second-order correction terms, the pairing density contribution is relatively the most important one.
 Furthermore, we performed fully self-consistent hybrid HFB calculations with the second-order Thomas-Fermi for the continuum
 and satisfactory results have been obtained. This will be particulary useful for the
 3D coordinate-space HFB calculations which are computationally very expensive.
  The second-order superfluid Thomas-Fermi method can be
 further extended by including the temperature dependence and rotational or vector fields.
 More interesting applications of the generalized
second-order Thomas-Fermi approximation are intended for large complex inhomogeneous superfluid systems such as
 cold atomic condensates and neutron star crusts, in which the pairing fields are very large compared to finite nuclei.

\begin{acknowledgments}
 Useful discussions with W. Nazarewicz, F.R. Xu, G. Fann, N. Hinohara, R. Id Betan, Y. Shi are gratefully acknowledged.
 This work was supported by the National Natural Science Foundation of China under Grants No.11375016, 11522538, 11235001,
and by the Research Fund for
the Doctoral Program of Higher Education of China (Grant
No. 20130001110001); and the Open Project Program of State Key Laboratory of Theoretical Physics,
Institute of Theoretical Physics, Chinese Academy of Sciences, China (Grant No. Y4KF041CJ1).
We also acknowledge that computations in this work were performed in the Tianhe-1A supercomputer
located in the Chinese National Supercomputer Center in Tianjin.
\end{acknowledgments}

\appendix*
\section{Second-order Expansion Terms}

The expansion coefficients of the second-order superfluid Thomas-Fermi approximation
to the normal density in Eq.~(\ref{ndensity}) are listed here:
\begin{equation}
\begin{split}
&\eta_1= -\frac{3h\Delta^2}{32E^5}\\
&\eta_2= \frac{(2h^2-\Delta^2)\Delta}{16E^5}\\
&\eta_3= \frac{2h\Delta^2-h^3}{32E^5}\\
&\eta_4= \frac{4h^2\Delta^2-\Delta^4}{16E^7} \\
&\eta_5= -\frac{(2h^2-3\Delta^2)h\Delta}{8E^7}\\
&\eta_6= -\frac{5h^2\Delta^2}{16E^7}\\
&\eta_7= -\frac{2h\Delta^3-3h^3\Delta}{16E^7}\\
&\eta_8= -\frac{3h^2\Delta^2-h^4-\Delta^4}{8E^7}\\
&\eta_9= -\frac{2h\Delta^3-3h\Delta^3}{16E^7}\\
&\eta_{10}= \frac{h^3+h\Delta^2}{16E^7}\\
\end{split}
\end{equation}

The expansion coefficients of the second-order superfluid Thomas-Fermi approximation
to the pairing density in Eq.~(\ref{pdensity}):
\begin{equation}
\begin{split}
&\theta_1= -\frac{(\Delta^2-2h^2)\Delta}{32E^5}\\
&\theta_2= -\frac{h(h^2-2\Delta^2)}{16E^5}\\
&\theta_3= -\frac{3h^2\Delta}{32E^5}\\
&\theta_4=  \frac{(3\Delta^2-2h^2)h\Delta}{16E^7}\\
&\theta_5=  \frac{h^4+\Delta^4-3h^2\Delta^2}{8E^7}\\
&\theta_6= -\frac{(2\Delta^2-3h^2)h\Delta}{16E^7}\\
&\theta_7= -\frac{5h^2\Delta^2}{16E^7}\\
&\theta_8= -\frac{(2\Delta^2-3h^2)h\Delta}{32E^7}\\
&\theta_9= -\frac{h^4-4h^2\Delta^2}{16E^7}\\
&\theta_{10}=  \frac{\Delta}{16E^5}\\
\end{split}
\end{equation}

The non-zero derivative terms employed in Eq.~(\ref{ndensity}) and Eq.~(\ref{pdensity}):
\begin{subequations}
\begin{equation}
 h\overleftrightarrow{\Lambda}^2\Delta = \displaystyle f \nabla^2\Delta
\end{equation}
\begin{equation}
 h\overleftrightarrow{\Lambda}^2h=\displaystyle fp^2\nabla^2f+2f\nabla^2U-\frac{2}{3}p^2(\nabla f)^2\\
\end{equation}
\begin{equation}
(\Delta\overleftrightarrow{\Lambda}h)^2 = \displaystyle\frac{1}{3} f^2p^2(\nabla\Delta)^2
\end{equation}
\begin{equation}
\begin{array}{lcl}
h(\overleftrightarrow{\Lambda}h)^2&=&\displaystyle-\frac{1}{12}fp^4(\nabla f)^2+f(\nabla U)^2+\frac{1}{6}f^2p^4\nabla^2f\vspace{5pt}\\
&&\displaystyle+\frac{1}{3}f^2p^2\nabla^2U+\frac{1}{3}fp^2(\nabla f)(\nabla U) \\
\end{array}\\
\end{equation}
\begin{equation}
 h(\overleftrightarrow{\Lambda} h)(\overleftrightarrow{\Lambda} \Delta)=\displaystyle\frac{1}{6}fp^2(\nabla f)(\nabla \Delta)+f(\nabla U)(\nabla \Delta)
\end{equation}
\begin{equation}
\Delta(\overleftrightarrow{\Lambda}h)^2 = \displaystyle \frac{1}{3}f^2p^2(\nabla^2\Delta)
\end{equation}
\begin{equation}
 h(\overleftrightarrow{\Lambda}\Delta)^2 = \displaystyle f (\nabla\Delta)^2
\end{equation}
\end{subequations}

\vspace{35pt}

\nocite{*}

%\bibliography{yinu}% Produces the bibliography via BibTeX.

\end{document}